\documentclass[12pt]{article}

\catcode`\@=11

\global\arraycolsep=2pt
\usepackage{amsbsy,amssymb,latexsym,amsfonts,amsmath}
\usepackage{graphicx,color}

\begin{document}
\begin{flushright}
\parbox{4.2cm}
{IPMU12-0160}
\end{flushright}

\vspace*{0.7cm}

\begin{center}
{ \Large Supercurrent, Supervirial and Superimprovement}
\vspace*{1.5cm}\\
{Yu Nakayama}
\end{center}
\vspace*{1.0cm}
\begin{center}
{\it Kavli Institute for the Physics and Mathematics of the Universe,  \\ Todai Institutes for Advanced Study,
University of Tokyo, \\ 
5-1-5 Kashiwanoha, Kashiwa, Chiba 277-8583, Japan}
\vspace{3.8cm}
\end{center}

\begin{abstract} Supersymmetric field theories possess a rich structure in their supercurrent supermultiplets. Some symmetries are manifest in one supercurrent supermultiplet but not in the others; for instance, R-symmetry is manifest in the  R-multiplet but not in the Ferrara-Zumino multiplet. Similarly, we argue that dilatation symmetry is manifest in the Virial multiplet (also known as a variant minimal supercurrent supermultiplet in the literature). It reveals that R-symmetry and dilatation symmetry are conceptually independent without further assumptions even though the superconformal symmetry connects the two. We show the structure of the Virial multiplet for general renormalizable supersymmetric field theories in $(1+3)$ dimension to all orders in perturbation theory, and discuss the condition for the dilatation invariance (but not necessarily R-symmetric nor superconformal). We present novel scale invariant trajectories with a nilpotent structure in coupling constants for non-unitary Wess-Zumino models with non-vanishing beta functions, which are, therefore, not superconformal.
\end{abstract}

\thispagestyle{empty} 

\setcounter{page}{0}

\newpage

\section{Introduction}

Symmetries play significant roles in physics. In particular, the dilatation invariance, or scale invariance appears everywhere. We can explain an intuitive reason why the scale invariance is ubiquitous from the philosophy of the renormalization group. In the long distance, or  far infrared, we forget the detailed information of the ultraviolet physics, and after integrating out the microscopic degrees of freedom, the universal infrared physics, associated with the renormalization group fixed point, emerges. The renormalization group fixed point suggests the invariance under scale transformation by its definition.\footnote{This does not explain why the scale invariant fixed point must exist. In relativistic systems, we believe Zamolodchikov's c-theorem and their generalizations in higher dimension will answer this question in a positive way.}

In relativistic quantum field theories, it has been observed that in almost all cases we know, the scale invariant fixed points show an enhanced symmetry known as conformal invariance. It turns out that the conformal symmetry is the largest bosonic space-time symmetry which is compatible with the existence of a non-trivial relativistic S-matrix \cite{Haag:1974qh}. We note that without relativity, there are many examples of scale invariant phenomena with no enhanced (conformal) symmetry. For example, we cannot even imagine what would be the conformal extension of the Lifshitz scale invariance ($t \to \lambda^Z t, x_i \to \lambda x_i$). Another example is the Efimov effect \cite{Efimov:1970zz}, in which the renormalization group flow is cyclic.

Why does the scale invariance imply conformal invariance? In (1+1) dimensional relativistic quantum field theories, Zamolodchikov and Polchinski gave a beautiful answer to this question \cite{Zamolodchikov:1986gt}\cite{Polchinski:1987dy} (see also \cite{Mack1}). Their theorem tells that  when the theory is (1) unitary, (2) is Poincar\'e invariant (in particular causal) and (3) has a discrete spectrum, then it must be conformal invariant.

The generalization of the theorem  in higher dimensions is not trivial at all \cite{Dorigoni:2009ra}. First of all, we do have counterexamples in space-time dimension $d>4$ \cite{Jackiw:2011vz}\cite{ElShowk:2011gz}. One may circumvent these examples by adding extra assumption: ``Noether assumption" which requires that the physically well-defined scale invariant current (rather than charge) must exist. Even then, while numerous studies both from field theories \cite{Nakayama:2011tk}\cite{Nakayama:2011wq}\cite{Luty:2012ww} and holographic analysis indicate the enhanced conformal invariance \cite{Nakayama:2009fe}\cite{Nakayama:2009qu}\cite{Nakayama:2010wx}\cite{Nakayama:2010zz}, the jury is still out, in particular, in $(1+3)$ dimension (c.f. \cite{Fortin:2012cq}\cite{Fortin:2012hn} for discussions in favor of the existence of scale invariance without conformal invariance: we will make a comment in Appendix C).

The conformal invariance as well as superconformal invariance gives a strong constraint on physical observables \cite{Mack:1975je}. The unitarity constraint for the both cases has been well-studied in the literature (see e.g. \cite{Minwalla:1997ka} for a review in various dimensions). The arguments are based on the unitary representations of the (super)conformal charges as quantum mechanical Hermitian operators, so the results are robust and universal. It is possible, however, that the existence of current, based on the above-mentioned Noether assumption, gives a stronger constraint. A canonical example would be the enhancement of scale invariance to conformal invariance, or Zamolodchikov's c-theorem \cite{Zamolodchikov:1986gt}. The statement of the c-theorem makes sense only if the energy-momentum tensor exists!

 Supersymmetric field theories possess a rich structure in their supercurrent supermultiplets. This is because there are some ambiguities in choosing the supersymmetric completion of the multiplets containing the energy-momentum tensor and the supercurrent (see e.g. \cite{Komargodski:2010rb}\cite{Dumitrescu:2011iu} and references therein for renewed interest in this subject). The most famous supercurrent supermultiplet is the Ferrara-Zumino supermultiplet \cite{Ferrara:1974pz}, but neither R-symmetry nor dilatation symmetry are manifest in this multiplet. Given our interest in scale invariance (possibly without conformal invariance), it is desirable to construct the supercurrent supermultiplet which shows manifest dilatation invariance as the existence of the R-multiplet \cite{Gates:1981yc} shows the manifest R-invariance.

We propose such a multiplet --- dubbed ``Virial multiplet"  in this paper. Actually, the Virial multiplet was known in the literature as a variant minimal supercurrent supermultiplet \cite{Kuzenko:2010am}\cite{Kuzenko:2010ni}, but the structure and the consequence of its existence have  not been studied very much (see e.g. \cite{Zheng:2010xx} for an attempt). We will clarify the relationship between the existence of the  Virial multiplet and the condition for the scale invariance (without superconformal invariance), which was also studied in \cite{Antoniadis:2011gn} from a different approach. The main difference is that they assumed the extra R-symmetry, but we do not. The existence of {\it both} Virial multiplet and R-multiplet leads to the same conclusion as theirs. We also study the most generic superimprovement possible, as far as we know, for each supermultiplets.

The outline of the paper is as follows. In section 2, we first review the space-time symmetry and the structure of the energy-momentum tensor. The distinction between scale invariance and conformal invariance is emphasized. We review the known supercurrent supermultiplets such as Ferrara-Zumino multiplet, R-multiplet and S-multiplet. We mention the possible superimprovement for each multiplets.  In section 2.3, we introduce the Virial multiplet, which shows the manifest dilatation symmetry. In section 3, we discuss various examples of Virial multiplets. In section 3.1 we study the free two-form tensor theory, and in section 3.2 we study the most generic renormalizable supersymmetric gauge theories (without Fayet-Iliopoulos terms). In section 3.3, we show novel scale invariant trajectories with a nilpotent structure in coupling constants for non-unitary Wess-Zumino models with non-vanishing $\beta$ functions, which are, therefore, not superconformal. This paper includes two appendices. In appendix A, we discuss the ambiguities in renormalization group $\beta$ functions. In appendix B, we study the component expressions of the conditions on scale invariance in non-unitary Wess-Zumino models. We have added appendix C to argue that the formula in section 3 directly computes the trace of the energy-momentum tensor with no contributions from the $S$-function in supersymmetric field theories. We also make a clarification on the recent debate over scale invariant but non-conformal field theories.

\section{Supercurrent, Supervirial and Superimprovement}
The goal of this section is to introduce the Virial supercurrent supermultiplet that shows the manifest dilatation invariance in supersymmetric field theories in $(1+3)$ dimension. We first review how the dilatation symmetry is realized as a space-time symmetry and discuss the structure of the supercurrent supermultiplet. 

\subsection{Virial current and improvement of energy-momentum tensor}
We would like to begin with the review of properties of the energy-momentum tensor and the space-time symmetry. 
Any continuous symmetry of the physical system requires the existence of a conserved charge, which is realized by an Hermitian operator on the physical Hilbert space. This is a consequence of the fundamental principle of quantum mechanics. In the quantum field theory, we typically {\it assume} that these charges can be constructed out of local currents. This is obvious when the theory has a Lagrangian description\footnote{However, the Lagrangian may not be gauge invariant or physical, and in such a situation, although the current may exist, it may not be well-defined globally. We will discuss related issues in this paper.} due to Noether's theorem. 
 In non-Lagrangian theories, the argument is not so obvious, but we will assume this locality requirement, which we call the ``Noether assumption", in the most part of this paper. When there is no local energy-momentum tensor, the coupling to  (Einstein) gravity is highly non-trivial.

Under the Noether assumption, the Poincar\'e invariance, which we also assume in this paper, implies that we have a conserved energy momentum tensor
\begin{align}
\partial^\nu T_{\mu\nu} = 0 , \label{conserv}
\end{align}
generating the space-time translation $x^\mu \to a^\mu$. The Lorentz invariance, in particular, demands that the energy-momentum tensor is chosen to be symmetric: $T_{\mu\nu} = T_{\nu \mu}$ such that the conserved Lorentz current is represented by $J_\mu^{(M_{\rho\sigma})} = x_{\rho}T_{\mu \sigma} - x_{\sigma}T_{\mu \rho}$. In the most part of the paper, we assume that the energy-momentum tensor is chosen to be symmetric by suitable improvements (known as Belinfante improvement).

In addition to the Poincar\'e invariance, we may require the invariance under dilatation $x^\mu \to \lambda x^\mu$. Under the Noether assumption, the dilatation invariance, or scale invariance, requires that the trace of the energy-momentum is given by a divergence of a certain current $J^\mu$ (known as Virial current)
\begin{align}
T^{\mu}_{\ \mu} = \partial^\mu J_\mu \ . \label{defv}
\end{align}
so that the dilatation current $D_\mu = x^\nu T_{\nu\mu} - J_\mu$ is conserved. Although we can always assume $T_{\mu\nu}$ to be symmetric in Poincar\'e invariant field theories, we did not use the symmetric property in the construction of the dilatation current. The theory can be dilatation invariant without (manifest) Lorentz invariance. This will be important in later sections. The Virial current is not unique: we can add any conserved current to it and  it does not change \eqref{defv}.

The conformal invariance is the largest space-time bosonic symmetry of the S-matrix and requires the invariance under the special conformal transformation:
\begin{align}
 x^\mu \to \frac{x^\mu + v^\mu x^2}{1+ 2v^\mu x_\mu+ v^2 x^2} \ . \label{sctf}
\end{align}
 The condition for a relativistic theory to be conformal invariant is that there exists a energy-momentum tensor whose trace vanishes. Then we can construct the conserved current 
\begin{align}
K^{(\rho)}_\mu = \left(\rho_\nu x^2 - 2x_\nu (\rho_\sigma x^\sigma)\right)T^{\nu}_{\ \mu}
\end{align}
that would generate the special conformal transformation, where $\rho_\nu$ is an infinitesimal parameter for $v_\nu$ in \eqref{sctf}

The energy-momentum tensor is ambiguous. We can modify the energy-momentum tensor as we wish as long as the conservation property \eqref{conserv} is kept intact. Suppose we have already used some of the ambiguities to make it symmetric, which is always possible as shown by Belinfante. The remaining ambiguities typically take the following three forms without further assuming any special conditions on the fields that can appear in the improvement terms.

First of all, the scalar improvement is given by
\begin{align}
\tilde{T}_{\mu\nu} = 
T_{\mu\nu} + (\partial_\mu \partial_\nu - \eta_{\mu\nu} \partial^2) L \ . \label{sci}
\end{align}
for any unconstrained scalar operator $L$. This improvement is canonical in scalar field theories.

The rank-two tensor improvement is given by
\begin{align}
\tilde{T}_{\mu\nu} = 
T_{\mu\nu} + \frac{1}{2} \left(\partial_\mu \partial_\rho L^\rho_{\ \nu} + \partial_\nu \partial_\rho L^\rho_{\ \mu} - \partial^2 L_{\mu\nu}  - \eta_{\mu\nu} \partial_\rho \partial_\sigma L^{\rho\sigma}\right) \ . \label{ti}
\end{align}
for any unconstrained traceless  symmetric tensor $L_{\mu\nu}$. The improvement is less common than the scalar improvement, but we sometimes encounter it in (higher-rank-tensor) gauge theories as well as higher derivative theories.

The rank-four tensor improvement is given by
\begin{align}
\tilde{T}_{\mu\nu} = T_{\mu\nu} + \partial_\rho \partial_\sigma L^{\rho \sigma}_{\ \ \mu \nu} \  \label{fi}
\end{align}
for any unconstrained traceless rank-four tensor $L_{\mu\nu \rho\sigma}$, which has the same symmetry as the Weyl tensor. Note that this improvement does not change the trace of the energy-momentum tensor. As a consequence, if the theory admits the rank-four tensor improvement, the traceless energy-momentum tensor is not unique for a given theory in the flat space-time.\footnote{A similar situation happens in Liouville theory in $(1+1)$ dimension (see e.g. \cite{Nakayama:2004vk} for a review).}

These improvements have the curved space-time interpretation as adding the non-minimal gravity coupling 
\begin{align}
S_{c} = \int d^4x \sqrt{g} (RL + R_{\mu\nu} L^{\mu\nu} + R_{\mu\nu\rho\sigma} L^{\mu\nu\rho\sigma}) \label{gravic}
\end{align}
with $\delta T_{\mu\nu} = \frac{\delta S_c}{\delta g^{\mu\nu}} |_{g_{\mu\nu} = \eta_{\mu\nu}}$. Note that $L$, $L_{\mu\nu}$ and $L_{\mu\nu\rho\sigma}$ are unconstrained.

One may slightly broaden the allowed improvement terms (see e.g.  \cite{Dumitrescu:2011iu}) by using ``field strength" rather than ``potential". The scalar improvement for instance can be relaxed by
\begin{align}
T_{\mu\nu} \to T_{\mu\nu} + \partial_\mu B_\nu - \eta_{\mu\nu} \partial^\rho B_\rho
\end{align}
with $B_\mu$ satisfying the ``Bianchi identity" $\partial_{[\mu} B_{\nu]} = 0$.   The local solution $B_\mu = \partial_\mu L$ reduces to \eqref{sci}. In this paper, we will not get involved with this kind of ``local vs global" issue. We only point out that without the potential $L$ the coupling to the non-linear gravity such as \eqref{gravic} is quite non-trivial.

Given these improvements, let us state the precise condition for conformal invariance \cite{Coleman:1970je}. Under the Noether assumption, the necessary and sufficient condition is that the trace of the symmetric energy-momentum tensor takes the form
\begin{align}
T^\mu_{\ \mu} = \partial_\mu \partial_\nu L^{\mu\nu} \ .
\end{align}
Then one can improve the energy-momentum tensor to be traceless and the theory is conformal invariant. Note that we have not used unitarity nor causality assumption here.

We notice that the tensor improvement may not be very useful when we would like to determine whether a given scale invariant field theory can be conformally invariant with the assumption of unitarity. This is because the possible improvement terms must be essentially trivial due to the unitarity constraint from the scale invariance \cite{Grinstein:2008qk} (with the additional assumption of diagonalizability and reality of the dilatation spectrum: without these assumptions, the conclusions are less clear \cite{Fortin:2011bm}). We, however, encounter the necessity of tensor improvement in several interesting examples in later sections.

Finally, we would like to make a small comment on the conformally invariant field theories without conformal current (nor energy-momentum tensor). In $(1+3)$ dimension, the conformally covariant massless wave-equations (known as  Bargmann-Wigner equation) must take the form \cite{Bracken:1982ny}
\begin{align}
\partial^{\alpha \dot{\beta}} \Psi_{\alpha \beta \gamma \cdots} = 0 \ , \label{wave}
\end{align}
where $\Psi_{\alpha \beta \gamma \cdots}$ has completely symmetric spinor indices. The CPT conjugate equation is obtained with dotted spinors. We note that except for helicity $0,1/2,1$, there is no simple Lagrangian formulation whose equations of motion are equivalent to \eqref{wave} with the local energy-momentum tensor. After the second quantization, the theory is conformal invariant in the sense that all the correlation functions transform in a conformally covariant manner as the helicity $h$ field $\Psi_{\alpha \beta \gamma \cdots}$  being conformal primaries with conformal dimension $h+1$.

An interesting observation is that the massless rank-four spinor $\Psi_{\alpha\beta\gamma \delta}$, which has helicity two, can be regarded as the linearized Weyl tensor around the Minkowski vacuum. Indeed, the linearized Einstein equation makes  the other components of linearized curvature tensor vanish, and the Bianchi identity is nothing but the conformal wave equation \eqref{wave}. Therefore, the linearized Einstein gravity {\it is} on-shell conformal invariant. However, there is no conserved energy-momentum tensor, nor conformal current, so the Noether assumption is clearly violated \cite{Dorigoni:2009ra}.\footnote{The vacuum Einstein equation $R_{\mu\nu} = 0$ is not Weyl invariant, so the Minkowski vacuum after Weyl transformation would not solve the vacuum Einstein equation. The conformal transformation here only acts on the linearized variation from the Minkowski vacuum. We also note that $\Psi_{\alpha \beta \gamma \delta}$ has the conformal dimension $3$, which may be unexpected from the non-linear Einstein gravity.} 

The discussion also applies to helicity 3/2 massless rank-three spinor $\Psi_{\alpha\beta\gamma}$ with the conformal wave-equation \eqref{wave}. The theory is equivalent to the on-shell Rarita-Schwinger theory of a massless spin 3/2 particle. Combining it with the above helicity 2 wave-equation, we can conclude that the linearized supergravity around the Minkowski vacuum is on-shell superconformal invariant. Again, we do not have a conserved supercurrent supermultiplet nor a superconformal current.\footnote{To some extent, this is just a peculiar coincidence in $(1+3)$ dimension. In higher dimensions, not all massless equations (like Maxwell theory or linearized gravity) are conformal invariant \cite{Siegel:1988gd}.}

\subsection{Ferrara-Zumino multiplet, R-multiplet and S-multiplet}
The main focus of this paper is the structure of the supercurrent supermultiplet in $\mathcal{N}=1$ supersymmetric field theories in $(1+3)$ dimension. We start with the superconformal algebra, which is the maximal space-time extension of the $\mathcal{N}=1$ supersymmetry with a non-trivial S-matrix. 
We begin with the conformal algebra
\begin{align}
i[M_{\mu\nu},P_\rho] &= \eta_{\mu\rho}P_\nu - \eta_{\nu\rho}P_\mu \ ,
\ \ i[M_{\mu\nu},K_\rho] = \eta_{\mu\rho}K_\nu - \eta_{\nu\rho}K_\mu \cr
i[M_{\mu\nu},M_{\rho\sigma}] &= \eta_{\mu\rho}M_{\nu\sigma} \pm \text{perm} \ ,  \ \  i[D,M_{\mu\nu}]  =0  \ , \ \
i[D,K_{\mu}] = - K_\mu \cr
i[D,P_\mu]  &= P_\mu \ , \ \ 
i[P_\mu, K_\nu] = -2M_{\mu\nu} + \eta_{\mu\nu} D \ ,
\end{align}
where Poincar\'e symmetry + dilatation symmetry is a proper subalgebra. The $\mathcal{N} = 1$ superconformal algebra adds
\begin{align}
i[D,Q^\alpha] = \frac{1}{2} Q^\alpha \ , \ \ i[D,S^\alpha] = -\frac{1}{2} S^\alpha \cr
i[R,Q^\alpha] = Q^\alpha \ , \ \ i[R,S^\alpha] = -S^\alpha \cr
i[K^{\alpha \dot{\alpha}}, Q^{\beta}] = -S^{\dot{\alpha}} \epsilon^{\alpha \beta} \ , \ \  i[P^{\alpha \dot{\alpha}}, S^{\beta}] = -Q^{\dot{\alpha}} \epsilon^{\alpha \beta} \cr
\{Q^\alpha , Q^{\dot{\alpha}} \} = 2P^{\alpha \dot{\alpha}} \ , \ \ \{S^\alpha , S^{\dot{\alpha}} \} = 2K^{\alpha \dot{\alpha}} \cr
\{Q^\alpha , S^\beta \} = M^{\alpha \beta} + (D-\frac{3}{2}R) \epsilon^{\alpha \beta} \ . 
\end{align}

There are two distinctive subalgebras here. The one is supersymmetry + R-symmetry and the other is supersymmetry + dilatation symmetry. The study of the supercurrent structure of the supersymmetry with R-symmetry {\it and} scale invariance (but without superconformal invariance) was done in \cite{Antoniadis:2011gn}, and we would like to understand the case when there is no R-symmetry. Without the R-symmetry, the theory cannot be superconformal invariant.\footnote{This means that the most of our work would be moot if we were able to show that the scale invariance must imply the conformal invariance. Although with some illuminating works, the question is still open. Of course, without unitarity, we can provide many examples of scale invariance without conformal invariance as we will see some of them in this paper.}

The simplest and canonical supercurrent supermultiplet is given by the so-called Ferrara-Zumino multiplet \cite{Ferrara:1974pz}. We only need supersymmetry (without R-symmetry, dilatation symmetry, or superconformal symmetry) for this multiplet to exist. It is given by 
\begin{align}
\bar{D}^{\dot{\alpha}}J^{FZ}_{\alpha \dot{\alpha}} &= D_{\alpha} X  \cr
\bar{D}_{\dot{\alpha}} X &= 0 \ ,
\end{align}
where $J^{FZ}_{\alpha \dot{\alpha}} = -2\sigma^{\mu}_{\alpha \dot{\alpha}} J^{FZ}_{\mu}$ is a real vector superfield (when the theory is unitary).

This multiplet is important because the theory with this supermultiplet can be naturally coupled to the old minimal supergravity. Moreover, all the renormalizable supersymmetric field theories (without Fayet-Iliopoulos terms) possess this supercurrent supermultiplet for their classical action. If we allow the gauge non-invariance of the Ferrara-Zumino multiplet, it even exists for the theories with Fayet-Iliopoulos terms.\footnote{This lack of gauge invariance eventually leads to the necessity of gauging of R-symmetry in supergravity with Fayet-Iliopoulos terms.}

When the theory possesses an R-symmetry, we can construct the R-current supermultiplet \cite{Gates:1981yc}. The supersymmetry algebra suggests that the R-current supermultiplet is  a supercurrent supermultiplet that contains the energy-momentum tensor.
The R-multiplet is defined by
\begin{align}
\bar{D}^{\dot{\alpha}}J^{R}_{\alpha \dot{\alpha}} &= \chi_{\alpha}  \cr
\bar{D}_{\dot{\alpha}} \chi_{\alpha} &= \bar{D} \bar{\chi} - D\chi = 0 \ ,
\end{align}
where $J^{R}_{\alpha \dot{\alpha}} = -2\sigma^{\mu}_{\alpha \dot{\alpha}} J^R_{\mu}$ is a real vector superfield (when the theory is unitary). The equation dictates that the bottom component vector in $J^{R}_{\mu}|_{\theta = \bar{\theta}=0}$ is a conserved R-current.
The constraint on $\chi_\alpha$ has a generic solution $\chi_{\alpha} = \frac{3}{2}\bar{D}^2 D_{\alpha} Y$ with a real superfield $Y$. The potential $Y$ always exists at least locally, and the theory with R-multiplet can be locally translated to the theory with a Ferrara-Zumino multiplet by introducing 
\begin{align}
J^{FZ}_{\alpha \dot{\alpha}} &= J^{R}_{\alpha \dot{\alpha}} - [D_{\alpha}, \bar{D}_{\dot{\alpha}}] Y \cr
X &= -\frac{1}{2} \bar{D}^2 Y \ . 
\end{align}
Note that the energy-momentum tensor in the Ferrara-Zumino multiplet and the one in the R-multiplet are different but related by the improvement.

In the recent literature, the most generic (under reasonable assumptions) supercurrent supermultiplet was proposed and it is called S-multiplet \cite{Komargodski:2010rb}\cite{Dumitrescu:2011iu}. The S-multiplet is defined by
\begin{align}
\bar{D}^{\dot{\alpha}}J^{S}_{\alpha \dot{\alpha}} = \chi_{\alpha} + \mathcal{Y}_\alpha  \cr
\bar{D}_{\dot{\alpha}} \chi_{\alpha} = \bar{D} \bar{\chi} - D\chi = 0 \cr
\bar{D}^2 \mathcal{Y}_\alpha = D_\alpha \mathcal{Y}_\beta + D_\beta \mathcal{Y}_\alpha = 0 \ .
\end{align}
The constraint on $\mathcal{Y}_\alpha$ can be solved locally as $\mathcal{Y}_\alpha = D_\alpha \mathcal{X}$ with a chiral superfield $\bar{D}_{\dot{\alpha}} \mathcal{X} = 0$ so that 
\begin{align}
\bar{D}^{\dot{\alpha}}J^{S}_{\alpha \dot{\alpha}} = \chi_{\alpha} + D_\alpha \mathcal{X} \ .
\end{align}
 Furthermore when the constraint on $\chi_\alpha$ can be solved as $\chi_{\alpha} = \frac{3}{2}\bar{D}^2 D_{\alpha} Y$, which is always locally possible, we can remove it by the superimprovement and the whole multiplet reduces to the Ferrara-Zumino multiplet: 
\begin{align}
J^{FZ}_{\alpha \dot{\alpha}} &= J^{S}_{\alpha \dot{\alpha}} - [D_{\alpha}, \bar{D}_{\dot{\alpha}}] Y \cr
X &= \mathcal{X}  -\frac{1}{2} \bar{D}^2 Y \ . 
\end{align}

In components, the S-multiplet has the structure
\begin{align}
J^S_\mu &= j_\mu^{(S)} + \theta\left(S_\mu-\frac{1}{\sqrt{2}}\sigma_\mu \bar{\psi} \right) + \bar{\theta} \left(\bar{S}_\mu + \frac{1}{\sqrt{2}}\bar{\sigma}_\mu \psi\right) + \frac{i}{2}\theta^2 \partial_\mu x^\dagger - \frac{i}{2}\bar{\theta}^2 \partial_\mu x \cr
&+(\theta \sigma^\nu \bar{\theta}) \left(2T_{\mu\nu} - \eta_{\mu\nu} Z + \frac{1}{2}\epsilon_{\mu\nu\rho\sigma}\left(\partial^\rho  j^{(S)\sigma} + F^{\rho \sigma}\right)\right) + \cdots \cr
\mathcal{X} &= x + \sqrt{2}\theta \psi + \theta^2 (Z + i \partial^\mu  j_\mu^{(S)}) + \cdots \cr
\chi_\alpha &= -i\lambda_{\alpha}  + (\delta_{\alpha}^{\beta}  D - 2i\sigma^\mu \bar{\sigma}^\nu F_{\mu\nu}) \theta_\beta + \theta^2 \sigma_{\alpha \dot{\alpha}}^\mu \partial_\mu  \bar{\lambda}^{\dot{\alpha}} + \cdots \ ,
\end{align}
where the supercurrent $S_\mu^{\alpha}$ and the symmetric energy-momentum tensor $T_{\mu\nu}$ are conserved. In addition, the constraint demands that the two-form $F_{\mu\nu}$ is closed, and $D$ and $\lambda$ satisfy the algebraic condition 
\begin{align}
 D &= -4T^{\mu}_{\ \mu} + 6Z \cr
\lambda & = -2i\sigma^\mu \bar{S}_\mu + 3i\sqrt{2}\psi
\end{align}
The component expressions for the Ferrara-Zumino multiplet and the R-multiplet are obtained by setting $\chi_\alpha = 0$ or $\mathcal{X}=0$. For instance, we find that in the R-multiplet, the bottom component is a conserved R-current. In the Ferrara-Zumino multiplet, $X$ contains the trace of the energy-momentum tensor, so $X$ is often called a supertrace multiplet.

Let us discuss the ambiguity, which we refer to as ``superimprovement", of the supercurrents. It is convenient to begin with the S-multiplet because the multiplet structure is more flexible and it accommodates wider classes of superimprovement. The simplest superimprovement is the scalar improvement \cite{Komargodski:2010rb}\cite{Dumitrescu:2011iu} 
\begin{align}
J^{S}_{\alpha \dot{\alpha}} &\to J^{S}_{\alpha \dot{\alpha}} + [D_{\alpha},\bar{D}_{\dot \alpha} ] U \cr
\mathcal{X} &\to \mathcal{X} + \frac{1}{2}\bar{D}^2 U \cr
\chi_{\alpha} &\to \chi_{\alpha} + \frac{3}{2}\bar{D}^2 D_{\alpha} U \ ,
\end{align}
where $U$ is any real scalar supermultiplet. For any $U$, the constraint equation is satisfied, and induces the improvement of the energy-momentum tensor $\delta T_{\mu\nu} = \frac{1}{2}(\partial_{\mu} \partial_\nu - \eta_{\mu\nu} \partial^2) U|_{\theta = \bar{\theta} = 0}$ corresponding to \eqref{sci}. In particular, this superimprovement changes the trace of the energy-momentum tensor by $-\frac{3}{2}\partial^2 U|_{\theta = \bar{\theta} = 0}$.

The second possibility is the so-called (super)Weyl improvement. Let $L_{\alpha \beta \gamma}$ be a chiral superfield (i.e. $\bar{D}_{\dot{\alpha}} L_{\alpha \beta \gamma} = 0$) with $\alpha,\beta,\gamma$ totally symmetric so that it has the same symmetry as the Weyl tensor supermultiplet. We can superimprove the S-multiplet by
\begin{align}
J^{S}_{\alpha \dot{\alpha}} &\to J^{S}_{\alpha \dot{\alpha}} + (\partial_{\dot{\alpha} \beta} D_\gamma L_{\alpha}^{\  \beta \gamma} + c.c.)\cr
\mathcal{X} &\to \mathcal{X} \cr
\chi_{\alpha} &\to \chi_{\alpha} , \label{fis}
\end{align}
which would not change the conservation of the S-multiplet. Note that this superimprovement does not change $\mathcal{X}$ and $\chi_{\alpha}$ so it will not modify the trace of the energy-momentum tensor $T^{\mu}_{\ \mu}$. Clearly, this improvement corresponds to the rank-four tensor improvement of the energy-momentum tensor \eqref{fi}.

Finally, we consider the following superimprovement. We introduce the real vector superfield $L_{\alpha \dot{\beta}}$ and 
\begin{align}
J^{S}_{\alpha \dot{\beta}}& \to  J^{S}_{\alpha \dot{\beta}} + 2D^2 \bar{D}^2 L_{\alpha \dot{\beta}} + 4i \partial_{\gamma \dot{\tau}} D^{\gamma} \bar{D}^{\dot{\tau}} L_{\alpha \dot{\beta}} \cr
&-4i\partial_{\gamma}^{\ \dot \tau} D^\gamma \bar{D}_{\dot{\beta}} L_{\alpha \dot{\tau}} - 4i \partial_{\alpha}^{\ \dot{\tau}} D^\gamma \bar{D}_{\dot{\tau}} L_{\gamma \dot{\beta}} 
-  8 \partial_{\alpha \dot{\beta}} \partial^{\gamma \dot{\tau}} L_{\gamma \dot{\tau}} \cr
\mathcal{X} &\to \mathcal{X} + \bar{D}^2 (\partial_{\alpha \dot{\beta}} L^{\alpha \dot{\beta}}) \cr
\chi_\alpha &\to \chi_{\alpha} \ . \label{tis}
\end{align}
This improvement modifies the energy-momentum tensor, in particular its trace changes as $\delta T^{\mu}_{\ \mu} = \Box \partial_\mu L^\mu|_{\theta = \bar{\theta} = 0}$. It is closely related to the rank-two tensor improvement \eqref{ti} with $L_{\mu\nu} = \partial_\mu L_\nu + \partial_\nu L_\mu$. 

Our search for the superimprovement is not systematic. In particular, it is not  clear to us whether there is no tensor improvement more generic than \eqref{tis}. We also note that generically, it may be possible that the superimprovement can be used to go beyond the S-multiplet by introducing higher-spin compensators in non-minimal supergravity.

If we combine some of the superimprovement of the S-multiplet and restrict the parameters, we can introduce the superimprovement within more constrained supercurrent supermultiplets. For the Ferrara-Zumino multiplet, $U$ appearing in the scalar superimprovement must be $U = \Xi + \bar{\Xi}$, where $\Xi$ is a chiral superfield. The improvement corresponding to \eqref{fis} and \eqref{tis} are not affected. On the other hand, for the R-multiplet, the scalar superimprovement must be a conserved current $U = J$ such that $D^2 J = \bar{D}^2 J=0$, which corresponds to the ambiguity of the R-current. 
The rank-four tensor improvement takes the same form as \eqref{tis}  while the vector superimprovement must accompany the simultaneous scalar superimprovement $ U = -(\partial_{\alpha \dot{\beta}} L^{\alpha \dot{\beta}})$ so that the constraint $\mathcal{X} = 0$ is intact.

As in the non-supersymmetric case, we have the supergravity interpretation of the superimprovement discussed above. Let us focus on the Ferrara-Zumino multiplet that corresponds to the old minimal supergravity. The case with the R-multiplet is essentially the same except that we have to use the new minimal supergravity. In the case of the S-multiplet, we may use the linearized supergravity presented in \cite{Komargodski:2010rb} due to the lack of the full non-linear supergravity construction.

Let $\mathcal{R}$, $G_{\alpha \dot{\alpha}}$ and $\mathcal{W}_{\alpha\beta\gamma}$ be three supercurvature superfields in old minimal supergravity. Roughly speaking, $\mathcal{R}$ contains the Ricci scalar, and $G_{\alpha \dot{\alpha}}$ contains the Ricci tensor and  $\mathcal{W}_{\alpha\beta\gamma}$ contains the Weyl tensor in components. 

We will consider the non-minimal couplings in the old minimal supergravity:
\begin{align}
\int d^4x d^2 \theta E \left( \mathcal{R} L + (\bar{\mathcal{D}}^2 - 8R) G_{\alpha \dot{\alpha}} L^{\alpha \dot{\alpha}} + \mathcal{W}_{\alpha\beta\gamma} L^{\alpha\beta \gamma} \right) + c.c. 
\end{align}
We can compute the supercurrent as well as the supertrace from this action, and we see that the each term corresponds to the superimprovement discussed in this section for Ferrara-Zumino multiplets.

\subsection{Virial multiplet}
Now we would like to introduce the supercurrent supermultiplet in which the dilatation symmetry is manifest. The main observation is that the ``imaginary" partner of the divergence of the R-current is the trace of the energy-momentum tensor. To exploit the idea, we put $i$ in the definition of the R-current supermultiplet.

We define the Virial multiplet by
\begin{align}
\bar{D}^{\dot{\alpha}}J^V_{\alpha \dot{\alpha}} &= i \eta_{\dot{\alpha}} \cr
\bar{D}_{\dot{\alpha}} \eta_{\alpha} &= D \eta - \bar{D} \bar{\eta} = 0 \ .
\end{align}
The difference from the R-multiplet is $i$ in the right hand side of the first equation.
When we expand $J^V$ in component, we find (see \cite{Dienes:2009td} for useful component expressions)
\begin{align}
J_\mu^V &= j_\mu + \theta (S_\mu + \frac{1}{3}\sigma_\mu \bar{\sigma}^\nu S_\nu) + \bar{\theta}  (\bar{S}_\mu + \frac{1}{3}\bar{\sigma}_\mu {\sigma}^\nu \bar{S}_\nu) + 2(\theta \sigma^\nu \bar{\theta}) \hat{T}_{\mu\nu} + \cdots \cr
\eta_\alpha &=-i\lambda'_{\alpha} + (\delta_{\alpha}^{\beta} D' - 2i \sigma^{\mu} \bar{\sigma}^\nu F'_{\mu\nu})\theta_\beta + \theta^2 \sigma_{\alpha \dot{\alpha}}^\mu \partial \bar{\lambda}^{'\dot{\alpha}} + \cdots \ , 
\end{align}
where
\begin{align}
\hat{T}_{\mu\nu} &= T_{\mu\nu} - \frac{1}{4}F'_{\mu\nu} + \frac{1}{2}\epsilon_{\mu\nu\rho\sigma} (\partial^\rho j^\sigma - \partial^\sigma j^\rho) \cr
D' & = -\partial^\mu j_\mu \ \ \ \lambda'_\alpha = \frac{1}{3}(\sigma_{\alpha \dot{\alpha}}^\mu \bar{S}_\mu^{\dot{\alpha}}) \ . 
\end{align}
In these expressions, $F'_{\mu\nu}$ is antisymmetric, and closed $dF' = 0$.  The tensor $T_{\mu\nu}$ is ``energy-momentum tensor" in the sense that it is conserved $\partial^\nu T_{\mu\nu} = 0$ so the theory is translationally invariant. In addition, the constraint shows that it is 
traceless $T^{\mu}_{\ \mu} = 0$ as expected from the fact that the constraint on the R-multiplet make the divergence of the R-current vanish, and here, its ``imaginary partner" vanish.  
It turns out, however, this ``energy-momentum tensor" is  not symmetric \cite{Dumitrescu:2011iu} (see also \cite{Zheng:2010xx}): $T_{\mu\nu} -T_{\nu\mu} = \frac{1}{4}F'_{\mu\nu}$.

Since the ``energy-momentum tensor" is traceless, theories with a Virial multiplet are scale invariant. As we have reviewed in section 2, this is because we can always construct the conserved dilatation current
\begin{align}
D_{\mu} = x^{\nu} T_{\nu\mu}
\end{align}
so that $\partial^\mu D_\mu = 0$ as long as  $\partial^\mu T_{\nu\mu} = 0$ and $T^{\mu}_{\ \mu} = 0$ irrespective of the antisymmetric part.
We would like to emphasize, however, we cannot conclude that the theory is conformal invariant even though the trace vanishes. This is because for the conformal invariance, we additionally need the ``symmetric" energy-momentum tensor. Furthermore, the bottom component of the Virial multiplet $j_\mu$ is not conserved (unless $\eta_\alpha = 0$) so the R-symmetry is not manifest. We recall that for the superconformal invariance, which is implied by the conformal invariance and supersymmetry, the R-symmetry is necessary.

 Since the energy-momentum tensor is not symmetric, the theory is not manifestly Lorentz invariant, which may sound fatal for the supercurrent supermultiplets for (super) Poincar\'e invariant field theories \cite{Dumitrescu:2011iu}. We, however, recall that the two-form $F'$ is closed. This means that we may introduce the potential $B_{\mu}$ by $\frac{-1}{4}F'_{\mu\nu} = \partial_\mu B_\nu - \partial_\nu B_\mu$. Then we can always improve the energy-momentum tensor
\begin{align}
\tilde{T}_{\mu\nu} = T_{\mu\nu} + \partial_\mu B_\nu - \eta_{\mu\nu} (\partial_\rho B^\rho) \ .
\end{align}
This improved energy-momentum tensor is still conserved $\partial^\mu \tilde{T}_{\mu\nu}$, and furthermore it is symmetric $\tilde{T}_{\mu\nu} = \tilde{T}_{\nu\mu}$. The price we paid is that it is no-longer traceless:
\begin{align}
\tilde{T}^{\mu}_{\ \mu} = -3\partial_\mu B^\mu \ .
\end{align}
This equation shows that $-3B_{\mu}$ is nothing but the Virial current, and the theory with a Virial multiplet is scale invariant but not necessarily conformal invariant.\footnote{The potential $B_{\mu}$ has a ``gauge symmetry" $B_\mu \to B_{\mu} + \partial_\mu \Lambda$, but this only adds $-3\partial^2 \Lambda$ to the trace of the energy-momentum tensor that can be removed by an improvement. We stress that the non-invariance of the energy-momentum tensor under this ``gauge symmetry" is not an issue rather it is a feature of the theory. It only means that there exists a dimension two scalar operator and the Virial current is not uniquely specified due to the improvement ambiguity. }

In the superfield formulation, the above discussion is equivalent to write (locally) $\eta_{\alpha} = -\frac{1}{2}\bar{D}^2 D_{\alpha} \hat{O}$ with a real superfield $\hat{O}$. Then the superimprovement tells us that the theory possesses a Ferrara-Zumino multiplet with $X = -\frac{i}{2}\bar{D}^2 \hat{O}$ by defining
\begin{align}
J^{FZ}_{\alpha \dot{\alpha}} = J^{V}_{\alpha \dot{\alpha}} - i\left\{D_{\alpha}, \bar{D}_{\dot{\alpha}}\right\} \hat{O}
\end{align}
If the theory has an additional R-symmetry, this $\hat{O}$ coincides with the one introduced in  \cite{Antoniadis:2011gn} as we will see.

As in the other supercurrent supermultiplets, the Virial multiplet enjoys the superimprovement. The scalar improvement adds 
\begin{align}
J^{V}_{\alpha \dot{\alpha}} &\to
J^{V}_{\alpha \dot{\alpha}} + i\left\{D_{\alpha}, \bar{D}_{\dot{\alpha}}\right\} J \cr
\eta_{\alpha} &\to \eta_{\alpha} - \frac{1}{2}\bar{D}^2 D_{\alpha} J \ ,
\end{align}
where $D^2 J = \bar{D}^2 J = 0$ is the conserved current superfield. This corresponds to adding a conserved current to the Virial current. The rank-four tensor improvement is also trivially applied. 

Let us summarize the condition for the existence of the Virial multiplet (and hence the condition for the dilatation invariance) for given supersymmetric field theories with a Ferrara-Zumino multiplet. When
\begin{align}
 X = \bar{D}^2 \left(\bar{\Xi} + \partial^\mu L_\mu -\frac{1}{2} i\hat{O} \right)  \label{FV}
\end{align}
for a certain antichiral $\bar{\Xi}$, real $L_\mu$ and real $\hat{O}$, 
we can superimprove the supercurrent and the Virial multiplet with $\eta_{\alpha} =  -\frac{1}{2}\bar{D}^2 D_{\alpha} \hat{O} $ exists (up to the above-mentioned ambiguities within the Virial multiplet). The first two terms in \eqref{FV} can be removed within the Ferrara-Zumino multiplet as discussed in the previous subsection.

On the other hand, the R-invariance requires that the Ferrara-Zumino supermultiplet has the structure
\begin{align}
 X = \bar{D}^2 \left(\bar{\Xi} + \partial^\mu L_\mu - \frac{1}{2}U\right)
\end{align}
for a certain antichiral $\bar{\Xi}$, real $L_\mu$ and real $U$. Then one can construct the R-multiplet with $\chi_{\alpha} = \frac{3}{2} \bar{D}^2 D_\alpha U$.  Therefore, the dilatation invariance and R-invariance are conceptually completely independent. We can even satisfy the both simultaneously without superconformal invariance. It requires 
\begin{align}
\bar{D}^2 (U -i\hat{O}) = 0 \ .
\end{align}
This is equivalent to the conclusion in \cite{Antoniadis:2011gn}, where we need two real scalar singlet superfields with dimension two in R-symmetric scale invariant supersymmetric field theories without superconformal invariance. 

We stress without R-symmetry, the theory can still be dilatation invariant. We give one (probably unphysical) example of such a theory. Consider the supersymmetric field theory for a chiral superfield $\Phi$ with the classical Lagrangian
\begin{align}
L = \int d^4 \theta (\Phi^2 + \bar{\Phi}^2)|\Phi|^2 + \int d^2\theta \Phi^6 + \int d^2\bar{\theta} \bar{\Phi}^6 \ . 
\end{align}
By assigning the dilatation charge $1/2$ to $\Phi$, the theory is classically dilatation invariant, but it is not R-symmetric (the superpotential is added to prevent the R-charge assignment zero to $\Phi$). The dilatation invariant vacuum $\Phi = 0$ is singular, so the quantization (e.g. unitarity) is non-trivial (see  \cite{Jackiw:2011vz}\cite{Polchinski:1987dy} for a similar non-supersymmetric theory), and we have no intention to attempt the quantization here. In section 3.2, we argue that perturbative dilatation invariance implies the R-symmetry in general renormalizable supersymmetric field theories.

The superconformal invariance requires $\hat{O} = U = 0 $ up to superimprovements, and we can construct the Ferrara-Zumino multiplet with $X=0$. 
When the theory admits the (super)Weyl improvement, such Ferrara-Zumino multiplet is not unique. A different choice of $L_{\alpha\beta\gamma}$ will lead to a different superconformal field theory in the curved background.

The coupling of the Virial multiplet to supergravity is interesting and worth studying further in the future. We only know how to do it in the linearized supergravity \cite{Buchbinder:2002gh}\cite{Gates:2003cz}. Of course, we can always transform it to the Ferrara-Zumino multiplet when it exists, and we can couple it to the old minimal supergravity at the fully non-linear level. 
At the linearized level, gravitational degrees of freedom that can naturally couple to the Virial multiplet in Wess-Zumino gauge consist of one {\it traceless}  symmetric tensor, one antisymmetric tensor (with gauge symmetry) and  fermions. The usual trace mode of the graviton (before fixing diffeomorphism invariance) is traded by the antisymmetric tensor, which is dual to a scalar on shell. It reminds us of the unimodular (super)gravity, and it seems intriguing to see the connection.

\section{Examples}
In this section, we will provide some examples of Virial multiplets. In section 3.1, we consider the free two-form tensor multiplet in which dilatation as well as superconformal invariance are subtly realized. In section 3.2, we discuss the  structure of the generic supersymmetric gauge theories in $(1+3)$ dimension and show the condition of the dilatation invariance to all orders in perturbation theory. In section 3.3, we present novel scale invariant trajectories with a nilpotent structure in coupling constants for non-unitary Wess-Zumino models.

\subsection{Free two-form tensor multiplets}
Free two-form tensor gauge theory in $(1+3)$ dimension is an interesting theory, where dilatation invariance and conformal invariance are both subtly realized (see e.g. \cite{Pons:2009nb} for a study of various $p$-form field theories in relation to conformal invariance). The two-form tensor field $ B = B_{\mu\nu}dx^{\mu}dx^\nu$ has a gauge invariance $B \to B + d\Lambda$, where $\Lambda$ is a one-form, and the gauge invariant three-form field strength $H = dB$. The action is
\begin{align}
S = \int d^4x H_{\mu\nu\rho} H^{\mu\nu\rho} \ .
\end{align}
The theory is free and every correlation functions are Gaussian, so we expect that the theory is dilatation invariant. We can compute the canonical symmetric energy-momentum tensor and its trace:
\begin{align}
T_{\mu\nu}  &= 3H_{\mu \rho \sigma} H_{\nu}^{\ \rho \sigma} - \frac{1}{2} \eta_{\mu\nu} H_{\rho\sigma \tau} H^{\rho  \sigma \tau} \cr
T^{\mu}_{\ \mu} &= H_{\rho\sigma \tau} H^{\rho  \sigma \tau} \cr
 &= 3 \partial^\mu ( H_{\mu \sigma \tau} B^{\sigma \tau}) \ . 
\end{align}
We see that the trace is a divergence of the Virial current, and the theory is scale invariant as expected. However, note that the Virial current is not gauge invariant, so the dilatation current $D_\mu = x^\nu T_{\mu\nu} - J_\mu = x^\nu T_{\mu\nu} - 3 H_{\mu \sigma \tau} B^{\sigma \tau}$ is not a gauge invariant object. The dilatation charge $D = \int d^3x D_0$,  however, is gauge invariant. 

The conformal invariance is more subtle. There is no ``local" operator made out of $B_{\mu\nu}$ which improves the energy-momentum tensor so that it is traceless. However, the free two-form gauge theory is on-shell equivalent to a free scalar (with a shift symmetry). Therefore, we can unitarily embed the two-form theory into a free conformal field theory (i.e. a free scalar) with the identification $*H = d\phi$. Every correlation functions for $H$ are conformal invariant as $\phi$ being a primary operator with conformal dimension $1$.

We will consider the supersymmetric generalization of the two-form gauge theory \cite{Siegel:1979ai}. The on-shell degrees of freedom is one two-form gauge field, one real scalar and one Majorana fermion. The theory is described by a real superfield $G$ with the Bianchi-identity $D^2 G = \bar{D}^2 G = 0$. It can be solved by using the potential chiral spinor superfield $\psi_\alpha$ with $G = D^\alpha \psi_\alpha + \bar{D}_{\dot{\alpha}} \bar\psi^{\dot{\alpha}}$ with the gauge invariance $\psi_\alpha  \to \psi_\alpha + i\bar{D}^2 D_{\alpha}V$. The $\theta \bar{\theta}$ component of $G$ has the three-form field strength $H$. The simple gauge invariant action
\begin{align}
S = \int d^4x d^4\theta G^2 \label{gfree}
\end{align}
provides the free two-form gauge theory (together with a free real scalar and a free Majorana fermion). 

The free two-form tensor theory with the action \eqref{gfree} is R-invariant with zero R-charge for $G$, and it is also scale invariant (but the scale current is not gauge invariant) without (manifest) conformal invariance as we can infer from the above discussion.
The gauge invariant R-multiplet is given by
\begin{align}
J^{R}_{\alpha \dot{\alpha}} = D_{\dot{\alpha}}G D_{\alpha} G
\end{align}
with $\chi_{\alpha} = \bar{D}^2 D_{\alpha} G^2$ (see e.g. \cite{Buchbinder:1995uq}). As we have discussed in the last section, non-trivial $\chi$ suggests that the theory is not manifestly superconformal invariant. From the structure of the R-current, we also understand that the theory possesses the gauge invariant Ferrara-Zumino multiplet with $X = -\frac{1}{3}\bar{D}^2 G^2$.

Since the theory is free, we expect the dilatation invariance. With the supersymmetry, the dilatation invariance requires that there exists a Virial multiplet. Indeed, one can show that
\begin{align}
\bar{D}^2 G^2 &= -4(i\partial^\mu ({\psi} \sigma_\mu \bar{D}^2 \bar{\psi})) \cr
&-\frac{1}{2}(\bar{D}^2 (D^2(\psi \psi)- \bar{D}^2 (\bar{\psi}\bar{\psi})) ))
-(\bar{D}^2 ((D\psi)^2 - (\bar{D} \bar{\psi})^2 )) \  \label{ocs}
\end{align}  
with the help of the equations of motion $D^2\bar{D}_{\dot{\alpha}} G = \bar{D}^2 D_{\alpha} G = 0$.
The first total divergence term in \eqref{ocs} is removed by the vector improvement discussed in section 2.3. The second line will appear in the right hand side of the conservation equation of the Virial multiplet.
From the analysis of section 2.3, by superimproving the supercurrent, we can construct the Virial multiplet with
\begin{align}
\frac{i}{3}  \hat{O} = -\frac{1}{2} (D^2(\psi \psi)- \bar{D}^2 (\bar{\psi}\bar{\psi})) 
- ((D\psi)^2 - (\bar{D} \bar{\psi})^2) \ .
\end{align}

We have to emphasize that the Virial multiplet is not gauge invariant under $\psi \to \psi + i\bar{D}^2 D_{\alpha}V$. This is a typical feature of ``scale invariance" in two-form field theories in $(1+3)$ dimension. As in the non-supersymmetric case, the scale current is not gauge invariant but the charge is gauge invariant. Similarly, the conformal invariance is not manifest at all in this formulation because there is no local superimprovement that will make $i\hat{O}$ vanish, but again we can dualize $G$ into a chiral scalar multiplet $\Phi$ as  $ G \sim \Phi+\bar{\Phi}$, and the dual theory can be naturally and unitarily embedded (see \cite{ElShowk:2011gz} for a discussion on the concept of conformal embedding) in a free superconformal field theory with one free chiral superfield $\Phi$, in which $\hat{O} = 0$ after superimprovement.\footnote{Notice, however, the superimprovement used here does not preserve the shift symmetry.}



\subsection{General renormalizable field theories}
Let us consider the most general renormalizable supersymmetric gauge theories without Fayet-Iliopoulos terms. To all orders in perturbation theory within the power-counting renormalization scheme, we can propose the ``exact" form of the Ferrara-Zumino multiplet. In the following expression, we assume the renormalization scheme where the gauge coupling constant is holomorphic while the matter kinetic term is canonical. We also assume that the gauge as well as gravity anomaly is absent. Let $\Phi_i$ be matter chiral superfields and $W^\alpha$ gauge superfields and $W$ gauge invariant superpotential. The Ferrara-Zumino multiplet 
satisfies (c.f. \cite{Clark:1980dw}\cite{Kogan:1995mr}\cite{Leigh:1995ep}\cite{Babington:2005vu}. See also \cite{Yonekura:2010mc} for a related S-multiplet.)
\begin{align}
\bar{D}^{\dot{\alpha}} J_{\alpha \dot{\alpha}}^{FZ} = \bar{D}_\alpha X \ ,
\end{align}
where the supertrace multiplet $X$ takes the form\footnote{We have deliberately written $\gamma_{ij}$ inside $\bar{D}^2$ although there is no difference outside here. When we promote the coupling constants to superfields, this will become important.}
\begin{align}
X = \frac{4}{3}\left[3W - \Phi_i \frac{\partial W}{\partial \Phi_i} - \frac{3C(G) -\sum_r C(r)}{32 \pi^2} \mathrm{Tr}(W^\alpha W_{\alpha}) - \frac{1}{8}\ \bar{D}^2 ( \gamma_{ij} \Phi^\dagger_i e^{-2V} \Phi_j) \right]  \ , \label{fzgeneral}
\end{align}
where the summation over the gauge group is implicit.
Here $\gamma_{ij}$ is the anomalous dimension matrix, whose explicit form is only known perturbatively by loop expansions. In the expression \eqref{fzgeneral}, we must use the Konishi-equation \cite{Konishi:1983hf}
\begin{align}
\frac{1}{4} \bar{D}^2 (\Phi^\dagger_i e^{-2V} \Phi_j)  = \Phi_i \frac{\partial W}{\partial \Phi_j} + \sum_r \frac{C(r)}{16\pi^2} \delta_{ij} \mathrm{Tr}(W^\alpha W_{\alpha}) \ .  \label{konishi}
\end{align}
This for instance leads to the $\beta$ functions of the Yukawa-coupling constant
\begin{align}
\beta_{Y_{ijk}} = \frac{dY_{ijk}}{d \log \mu} = \gamma_{il} Y_{ljk} + \gamma_{jl} Y_{ilk} + \gamma_{kl} Y_{ijl} \label{betay}
\end{align}
as well as the $\beta$ functions of the gauge coupling constants
\begin{align}
\beta_g = \frac{d (1/g^2)}{d\log \mu} = \frac{3C(G) - \sum_r(1- \mathrm{Tr}(\gamma_{r})) C(r)}{16\pi^2} \label{betag}
\end{align}
for each gauge group by noticing $T^{\mu}_{\ \mu} = \beta^I \mathcal{O}_I$ up to improvement.
This $\beta$ function corresponds to the holomorphic renormalization scheme where the denominator of the NSVZ $\beta$ function \cite{Novikov:1983uc} does not show up \cite{ArkaniHamed:1997ut}.\footnote{Note that the use of the Konishi-equation is crucial. In a similar way, we cannot justify the use of the naive equations of motion in the evaluation of the divergence of the Virial current. The claim that the equations of motion are exact inside the energy-momentum tensor is incompatible with the supersymmetry.}

For completeness, let us present $J_{\alpha \dot{\alpha}}^{FZ}$ in a schematic form. We will omit the detailed discussions on the renormalization. 
\begin{align}
J_{\alpha \dot{\alpha}}^{FZ} = 2(D_{\alpha} \Phi_i) (\bar{D}_{\dot{\alpha}} \bar{\Phi}_i) - \frac{1}{g^2} \mathrm{Tr}W_{\alpha} \bar{W}_{\dot{\alpha}} - \frac{2}{3}[D_{\alpha}, \bar{D}_{\dot{\alpha}}] (\delta_{ij} + \frac{1}{2}\gamma_{ij}) \bar{\Phi}_i e^{-2V}\Phi_j \ .
\end{align}
At the classical level, we may introduce the Fayet-Iliopoulos terms by adding $-\frac{2\xi}{3}[D_\alpha, \bar{D}_{\dot{\alpha}}]V$ to $J_{\alpha \dot{\alpha}}^{FZ}$ and $-\frac{\xi}{3} \bar{D}^2 V$ to $X$ (see e.g. \cite{Arnold:2012yi} and references therein).

Strictly speaking, when the theory contains a possible candidate for Virial currents, the $\beta$ functions are ambiguous (see \cite{Osborn:1991gm} appendix of \cite{Nakayama:2012sn}). We have a tacit assumption that we have fixed the scheme (or gauge of the source terms) in the above expression \eqref{betay}. Or we may substitute the scheme independent $\mathcal{B}$ functions instead. It is crucial that our supersymmetric computations of the renormalization group equation computes $\mathcal{B}$ function directly.
Further details of this point  can be found in Appendix A and C.

Our claim is that the above formula for $X$ is exact to all orders in perturbation theory. The formula cannot be non-perturbatively exact at least in two ways. First of all, we have not included any non-perturbative corrections to the superpotential such as ADS superpotential \cite{Affleck:1983mk}. Secondly, to all orders in perturbation theory, the supercurrent supermultiplet does not mix with any other currents or chiral operators whose bare scaling dimension is different. However, this is not true in non-perturbative regime such as the free-magnetic phase of SQCD, where the unitarity is naively violated in the electric picture. In such situations, the supercurrent does deviate from the expression in \eqref{fzgeneral} by mixing with the non-perturbative (and possibly non-local) operators.

We would like to study the condition for the enhanced symmetry. Let us begin with the R-symmetry.
The condition for the R-invariance (i.e. existence of an R-multiplet) is
\begin{align}
3W - \Phi_i \frac{\partial W}{\partial \Phi_i} - \frac{\tilde{\gamma}_{ij}}{2}\Phi_i \frac{\partial W}{\partial \Phi_j} = 0 \label{r1}
\end{align}
for a certain Hermitian $\tilde{\gamma}_{ij}$ (not necessarily the same $\gamma_{ij}$ in the Ferrara-Zumino multiplet) 
and vanishing of the R-anomaly
\begin{align}
3C(G)-\sum_r (1- \mathrm{Tr} (\tilde{\gamma}_r)) C(r) = 0 \  \label{r2}
\end{align}
for each gauge group.
We note that the difference $ U = (\gamma_{ij} - \tilde{\gamma}_{ij}) \Phi^\dagger_i \Phi_j$ appears in the R-multiplet $\chi_{\alpha}=\frac{3}{2}\bar{D}^2D_{\alpha} U$. When $U$ vanishes, the chosen R-symmetry is the superconformal R-symmetry and the theory is superconformal. We may find non-unique candidates for $\tilde{\gamma}_{ij}$ that solves \eqref{r1} and \eqref{r2}, in which situation, the so-called $a$-maximization principle \cite{Intriligator:2003jj} serves as a guidance to find the superconformal R-symmetry. As far as we know, there is no proof, however, that $\tilde{\gamma}_{ij}$ selected by the $a$-maximization has a solution $\tilde{\gamma}_{ij} = \gamma_{ij}$ with respect to coupling constants.

On the other hand, the condition for the scale invariance (i.e. existence of a  Virial multiplet) is 
\begin{align}
3W - \Phi_i \frac{\partial W}{\partial \Phi_i} - \frac{{\gamma}_{ij}}{2} \Phi_i\frac{\partial W}{\partial \Phi_j}  = -\frac{iQ_{ij}}{2} \Phi_i \frac{\partial W}{\partial \Phi_j}  \label{condis}
\end{align}
and 
\begin{align}
3C(G)-\sum_r (1- \mathrm{Tr} ({\gamma}_r - iQ_r)) C(R) = 0 \ . \label{esse}
\end{align} 
 for each gauge group with a certain Hermitian matrix $Q$. The condition \eqref{esse} essentially tells that the Virial current must be non-anomalous and gauge $\beta$-functions must vanish because there is no other way to remove $i$ in the equation. This is in accord with the observation that the gauge $\beta$ function necessarily vanishes in a perturbative scale invariant fixed point \cite{Polchinski:1987dy}\cite{Nakayama:2011tk}. When these equations are satisfied, there exists a Virial multiplet with $\hat{O} = Q_{ij} \Phi^\dagger_i e^{-2V}\Phi_j$ with $\eta_{\alpha} = -\frac{1}{2}\bar{D}^2 D_{\alpha} \hat{O}$. 

If the theory is scale invariant and R-invariant, then  both conditions must be satisfied. This means that we have $\bar{D}^2 U = i \bar{D}^2 \hat{O}$ as we have discussed in section 2.3. As we will show, it turns out that this is always the case in perturbative fixed points in most general unitary renormalizable gauge field theories considered here.  In non-perturbative regime, we emphasize that there is no necessity to assume the existence of two gauge singlet scalar operators $U$ and $\hat{O}$ for the scale invariance: we needed $U$ for the R-invariance and $\hat{O}$ for the scale invariance, and they are independent with no further assumptions. 

The (super)conformal invariance requires that $U = \hat{O} = 0$ (up to superimprovements). This leads to the usual assumption that R-charges are proportional to the anomalous dimensions. However, we should note that there is no guarantee that we can find the corresponding coupling constants for a given $\gamma_{ij}$ (which typically happens when we hit the unitarity bound). Another possibility is that we may be able to find a  different solution for $\gamma_{ij}$ with non-zero $Q_{ij}$ than the one predicted from the R-charges.

Within the perturbation theory, we can say a little bit more. Suppose we found a non-trivial Virial multiplet so that the theory is  scale invariant but not superconformal invariant in perturbation theory with respect to a certain small (loop-counting) expansion parameter $\hbar$. The Virial multiplet is non-trivial beyond a certain order  in $\hbar$. We know that at order $\mathcal{O}(\hbar^0)$, the Virial current component $\eta_{\alpha}$ must vanish because all the power-counting renormalizable classically scale invariant theories in $(1+3)$ dimension are conformal invariant at the classical level. This means that the theory must possess the classical R-symmetry (as can be easily seen by assigning $R(\Phi) = 2/3$ to all chiral matter superfields). 

The Virial multiplet is either trivial at $\mathcal{O}(\hbar)$ or non-trivial. The explicit computation of the anomalous dimension matrix shows that the latter is impossible when the theories are unitary. The argument is based on the idea \cite{Polchinski:1987dy}\cite{Dorigoni:2009ra}. The equation \eqref{condis} in perturbation theory gives
\begin{align}
\gamma_{il} Y_{ljk} + \gamma_{jl} Y_{ilk} + \gamma_{kl} Y_{ijl} = i\left(Q_{il} Y_{ljk} + Q_{jl} Y_{ilk} + Q_{kl} Y_{ijl}\right)  \ . \label{qy}
\end{align}
At one-loop, the anomalous dimension takes the form $\gamma_{ij} = \frac{1}{32\pi^2}Y_{ikl}\bar{Y}_{jkl} - \frac{g^2}{8\pi^2}C(r)_{ij}$. Here $C(r)_{ij} = (r_A r_A)_{ij}$. By acting $\bar{Y}Q$ on the both sides of \eqref{qy} and contracting indices, we see that the right hand side is a pure imaginary number, but the left hand side is a real number from the explicit form of one-loop $\gamma_{ij}$. Thus, $QY$ must vanish, which means that the Virial multiplet can be improved to be trivial at this order.

Thus, the perturbative scale invariance requires that the fixed points (trajectories) look like superconformal at order $\mathcal{O}(\hbar)$. The superconformal invariance means that the theory must possess the R-symmetry at order $\mathcal{O}(\hbar)$. However, the R-symmetry is one-loop exact, and the one-loop requirement of the R-symmetry is identical to solving the same linear equation \eqref{r1} required by the full R-symmetry, so the existence of the one-loop R-symmetry guarantees the existence of the $\tilde{\gamma}_{ij}$ that satisfies \eqref{r1}. We conclude that the theory possesses R-symmetry as well as the R-current supermultiplet.

This does not lead to the statement that the perturbative fixed points must be superconformal beyond $\mathcal{O}(\hbar)$. There is no a priori guarantee that $\gamma_{ij} = \tilde{\gamma}_{ij}$ continues to have a solution with respect to the coupling constants at higher order. This happens when the $\tilde{\gamma}_{ij}$ hits the unitarity bound, or when the theory is scale invariant but not conformal invariant. The explicit form of the $\beta$ function with the gradient flow property within perturbation theory (up to certain order checked in the literature \cite{Freedman:1998rd}) seems to suggest that the latter possibility is excluded. It would be interesting to have better conceptual understanding of the problem.

\subsection{Non-unitary Wess-Zumino model with novel scale invariant trajectory}
As we have seen, in unitary $(1+3)$ dimensional gauge theories, it is difficult to construct a non-trivial Virial multiplet in perturbation theory. Relaxing the condition of the unitarity leads to a plethora of examples (see \cite{Riva:2005gd}\cite{Ho:2008nr} for some examples in $1+1$ dimension).
Let us consider the following particular Wess-Zumino model in $(1+3)$ dimension  with three chiral superfields $\Phi_1$, $\Phi_2$, and $S$ whose superpotential is 
\begin{align}
W &= Y_{ij} S\Phi^i \Phi^j \cr
\bar{W} & = \bar{Y}_{ij} \bar{S} \bar{\Phi}^i \bar{\Phi}^j \ .
\end{align}
Here $(i = 1,2)$ and $Y_{ij}$ and $\bar{Y}_{ij}$ are symmetric but treated independently.\footnote{Before going into the explicit construction of the non-unitary scale invariant trajectories and the Virial multiplet, let us mention one excuse for the lack of unitarity. The independent treatment of the holomorphic superpotential and anti-holomorphic superpotential certainly breaks the unitarity, but if we are interested in certain observables, e.g. a chiral ring structure, the result formally would not be affected. In this way, it is possible to extract the physics of non-scale invariant unitary field theories out of the scale invariant non-unitary field theories.}

The existence of  the Virial multiplet, or the scale invariance demands (see Appendix B for further explanations in component formulation)
\begin{align}
\bar{D}^2(\gamma_{SS} \bar{S} S + \gamma_{ij} \bar{\Phi}^i \Phi^j - iQ_{ij}\bar{\Phi}^i\Phi^j) &= 0 \cr
{D}^2(\gamma_{SS} \bar{S} S + \gamma_{ij} \bar{\Phi}^i \Phi^j +iQ_{ij}\bar{\Phi}^i\Phi^j) &= 0 \ . \label{vseq}
\end{align} 
The second equation would have been an automatic consequence of the first one in unitary theory by complex conjugation. Here, however, we assume both $\gamma_{ij}$ and $Q_{ij}$ are non-Hermitian and require that both of \eqref{vseq} must be satisfied.

The one-loop computation gives
\begin{align}
\gamma_{SS} = \frac{1}{32\pi^2}Y_{ij} \bar{Y}_{ji} \cr
\gamma_{ij}= \frac{1}{32\pi^2}Y_{il} \bar{Y}_{lj}  \ ,
\end{align}
and by using the matrix notation, the equations \eqref{vseq} become
\begin{align}
\mathrm{Tr}(Y\bar{Y}) Y + (Y\bar{Y})Y- i32\pi^2QY &= 0 \cr
\bar{Y} \mathrm{Tr}(Y\bar{Y})  + \bar{Y}(Y\bar{Y})+ i32\pi^2\bar{Y}Q &= 0 \ . \label{vsbq}
\end{align}
We look for a non-trivial solution in which at least one of $QY$ or $\bar{Y}Q$ is non-zero.

A particular solution is
\begin{align}
Y_0 = y_0 \begin{pmatrix}
1 & 1 \\
1 & 1 
\end{pmatrix} \ ,  \ \ 
\bar{Y}_0 = \bar{y}_0 \begin{pmatrix}
1 & 0 \\
0 & -1 
\end{pmatrix}
\end{align}
with
\begin{align}
-iQ &= \frac{1}{32\pi^2}(y_0\bar{y}_0)  \begin{pmatrix}
1 & -1 \\
1 & -1 
\end{pmatrix} 
\end{align} \ ,
in which $y_0$ and $\bar{y}_0$ are independent arbitrary complex parameters.
The renormalization group flow $\frac{dY}{d\log\mu} = -iQY$ and $\frac{d\bar{Y}}{d\log\mu} = i\bar{Y}Q$ gives
\begin{align}
Y&= Y_0 \cr
\bar{Y} &= \bar{Y}_0 -i \frac{1}{32\pi^2}y_0\bar{y}_0^2\log \mu \begin{pmatrix}
1 & -1 \\
-1 & 1 
\end{pmatrix} \ . \label{traj}
\end{align}
Along the renormalization flow, the $\beta$ functions stay constant, so 
the solution is consistent with the renormalization group evolution.

Furthermore, the nilpotency of the coupling constants
\begin{align}
Y \bar{Y} = y_0 \begin{pmatrix}
1 & -1 \\
1 & -1 
\end{pmatrix} \ ,  \ \ 
Y\bar{Y} Y = 0 \ , \ \
\bar{Y} Y \bar{Y}=y_0\bar{y}_0^2 \begin{pmatrix}
1 & -1 \\
-1 & 1 
\end{pmatrix}
\end{align}
make all the higher loop corrections vanishes. In this way, the renormalization group trajectory \eqref{traj} gives a non-unitary example of scale invariant but non-conformal supersymmetric field theories in $(1+3)$ dimension. In unitary field theories, the eigenvalues of $Q$ are real, so the trajectories, if any, have cyclic behaviors \cite{Fortin:2011sz}\cite{Fortin:2012ic} but it is not guaranteed in non-unitary examples. Our model also reveals that the diagonalizability of the dilatation operator can be lost when the theory is not conformal invariant.

\section{Discussions}
One of the main motivations of the paper is to clarify the role of the Virial current and its supermultiplet in the supersymmetric field theories. We have shown that the dilatation invariance and the existence of the Virial current are manifested in the Virial supermultiplet, which was the least studied minimal supercurrent supermultiplet in the literature.

Since it is the least studied supercurrent supermultiplet, we even do not know the existence of the non-linear supergravity multiplet that will naturally couple to it while the linearized superspace action is known \cite{Gates:2003cz}. Such a gravity only couples to dilatation invariant field theories, and it shares some common features with the unimodular gravity. The role of the extra anti-symmetric tensor is mysterious, and the existence will deviate it from the Weyl gravity (where the unitarity is lost) even though the gravitational degree of freedom is traceless and expected to be unimodular at the non-linear level.

Of course, within our investigation, we can always stick to the Ferrara-Zumino multiplet and the old minimal supergravity. It would be interesting to investigate whether this Noether assumption for the Lorentz current, so that the Virial multiplet is always superimproved to be a Ferrara-Zumino multiplet, always holds or not. If this assumption is not true, eventually we have to consider the most generic $(20+20)$ supercurrent supermultiplet proposed in the literature \cite{Osborn:1998qu}\cite{Magro:2001aj}\cite{Kuzenko:2010am}.

We have discussed the structure of the possible Virial multiplet for generic supersymmetric gauge theories within perturbation theory. Even within the  perturbation theory, it would be very interesting to complete the analysis on the local supersymmetric renormalization group and obtain the consistency conditions.\footnote{The author would like to thank J.~Erdmenger for pointing out the reference \cite{Grosse:2007au} and related discussions.}  We notice that some of the superimprovements discussed in this paper play an important role in controlling the local renormalization group flow and understanding the scheme dependence. Such analysis would give us a hint of the mechanism of the enhancement of the superconformal invariance from the mere scale invariance or R-invariance.\footnote{Again, the Fayet-Iliopoulos term is an exception. The existence is consistent with the R-invariance, but it will spoil the dilatation invariance and hence the superconformal invariance.}
 Beyond the perturbation theory, almost nothing is known except that the assumption of superconformal invariance seems to work in all examples that we are aware of. It is urgent to understand the underlying reason why this is the case.

The construction of the holographic dual with the Virial multiplet would be very interesting. We have studied the non-supersymmetric situations in \cite{Nakayama:2009fe}\cite{Nakayama:2009qu}\cite{Nakayama:2011zw}, where the violation of the (strict) null energy condition was necessarily introduced. It is important to know how the supersymmetry may give more constraints on the energy-condition in the bulk.
The geometric interpretation of the cyclic renormalization group was clarified in \cite{Nakayama:2012sn}, where it was shown that the introduction of the Virial current is equivalent to the gauging of the bulk matter fields.
 In our novel example of a nilpotent structure in the renormalization group, it is likely that we have to introduce the non-unitary representations (more specifically nilpotent representations) of the gauge group in the bulk.

Finally, if we put supersymmetric theories on a curved background, the supercurrent structure must change accordingly.\footnote{Another interesting possibility is to consider the anti-non-commutative superspace. It shows scale invariance without superconformal invariance \cite{SN}. The Lorentz symmetry is partly broken, so it is important to understand the structure of the supercurrent supermultiplet.} In particular, there exist further anomalous contributions to the supercurrent conservations such as conformal anomaly. The supersymmetric field theories on rigid curved backgrounds have attracted a lot of attention in relation to the localization and exact evaluations of various physical quantities. The discussions so far mainly use the R-multiplet since the R-symmetry is useful to compensate the non-trivial spinor transformation in the curved backgrounds \cite{Festuccia:2011ws}\cite{Dumitrescu:2012ha}, but it would be interesting to see whether the Virial multiplet can be as effective by using dilatation symmetry instead of R-symmetry.

\section*{Acknowledgements}
The author would like to thank R.~Jackiw and M.~Duff for comments on scale invariance and conformal invariance. 
He would like to acknowledge M.~Buican for email correspondence.
The work is supported by the World Premier International Research Center Initiative of MEXT of
Japan.

\appendix

\section{Ambiguities in $\beta$ function and unambiguous $\mathcal{B}$ function}
When the theory possesses spin 1 currents with scaling dimension  3 in $(1+3)$ dimension, in particular, when the theory has a non-trivial candidate for the Virial current, the $\beta$ functions associated with the renormalization group are ambiguous  \cite{Osborn:1991gm} due to the operator identity such as $\beta^I O_I = \partial^\mu J_\mu$ in addition to the usual scheme dependence via the change of the coordinates in coupling constant space. 
The trace of the energy-momentum tensor $T^\mu_{\ \mu} = \beta^I O_I + S\cdot \partial^\mu J_\mu $ is invariant under
\begin{align}
\beta^I\ &\to \beta^I + (\omega^g g)^I \cr
S &\to S - \omega  \ . \label{gauge}
\end{align}
In this way, the renormalization of the coupling constant can also be regarded as the renormalization of the background current.

To avoid the ambiguity, we can introduce the ``gauge invariant" $\mathcal{B}^I$ function
\begin{align}
\mathcal{B}^I &= \beta^I + (S^g g)^I \ . 
\end{align}
When we discuss the change of ``coupling constants" along the renormalization group flow, it is more convenient to fix the ``gauge" (or scheme).\footnote{The reason we call it gauge is that it is precisely the gauge transformation of the external source that induces the ambiguities along the renormalization group flow. In the holographic renormalization group flow, it is even more obvious \cite{Nakayama:2012sn} because the operator identity is realized by gauging the bulk matter fields, and the ambiguities here are nothing but the gauge transformations in the bulk.} The most convenient one that is implicitly employed in the rigid background computation, and the best suited one to compare with the perturbative flat-space-time Feynman diagram analysis is to set $S=0$.  The merit of this gauge is that we can state that the change of the coupling constants (in this gauge) is described by the renormalization group equation $\frac{dg^I}{d\log\mu} = \mathcal{B}^I(g)$. We also note that the trace of the energy momentum tensor in generic gauge
\begin{align}
T^{\mu}_{\ \mu} = \beta^IO_I + S \cdot \partial^\mu J_\mu
\end{align}
becomes $ T^{\mu}_{\ \mu} =\mathcal{B}^I O_I$ in this $S=0$ gauge so that the conformal invariance is equivalent to $\mathcal{B}^I = 0$ while scale invariance is $\beta^I \sim 0$.

In our supersymmetric example discussed in section 3.2, the ambiguity boils down to the use of the Konishi-anomaly equation. The one thing which was not discussed in \cite{Osborn:1991gm} is the anomalous contribution from the measure of the path integral in deriving the consistency relations. Here, the anomalous contribution is encoded in the Konishi-anomaly equation. The above argument remains true by using the anomalous operator identity rather than the naive one.
\section{Scale invariance condition in non-unitary Wess-Zumino model in components}
We would like to justify the scale invariance conditions for non-unitary field theories used in section 3.3.
First of all, since we abandon the reality of the action, the Hamiltonian is not  a Hermitian operator. As a consequence, the chiral superfield $\Phi$ and anti-chiral superfield $\bar{\Phi}$ cannot be related by a simple complex conjugation. Indeed, if $\Phi$ were a Heisenberg field i.e. $\Phi(t) = e^{-iHt}\Phi(0)e^{+iHt}$, its complex conjugate would be $\Phi(t)^\dagger = e^{-iH^\dagger t} \Phi(0)^\dagger e^{+iH^\dagger t}$, and it would have no simple relation to $e^{-iH t} \bar{\Phi}(0) e^{iH t}$ even if $\Phi(0)^\dagger = \bar{\Phi}(0)$. The energy-momentum tensor and Virial currents are also not real in a conventional sense.

To justify \eqref{vseq}, let us focus on the Yukawa interaction. The supersymmetry automatically completes the $\phi^4$ interaction.
The trace of the energy-momentum  tensor must be 
\begin{align}
T^{\mu}_{\ \mu} = \beta_{Y_{a,bc}}  \phi_a \psi_b \psi_c + \beta_{\bar{Y}_{a,bc}} \bar{\phi}_a \bar{\psi}_b\bar{\psi}_c  + \mathcal{O}({\phi}^4)\ .
\end{align}
Here $a,b,c$ corresponds to $S,i,j$ in the model discussed in section 3.3. Note that the classical part algebraically vanishes up to classical improvement terms irrespective of the non-reality of the action.
We do not demand $\beta_{Y_{a,bc}}^\dagger = \beta_{\bar{Y}_{a,bc}}$, which would have been true in unitary theories. 

The scale invariance requires that the trace must be a divergence of the Virial current
\begin{align}
T^{\mu}_{\ \mu} = \partial^\mu J_{\mu} \ , \label{vcon}
\end{align}
where $J_\mu$ is not real any more. We expand the Virial current as $J_{\mu} = Q_{ab} \psi^a \sigma^\mu \bar{\psi}^b + \text{bosons}$. If we imposed the reality of $J_\mu$, $Q_{ab}$ must have been Hermitian, but we do not impose the condition here: otherwise it would be inconsistent with the non-reality of $T^{\mu}_{\ \mu}$. 

Now by using the Dirac equations with the Yukawa interaction, the equation \eqref{vcon} demands
\begin{align}
\beta_{Y_{a,bc}}  \phi_a \psi_b \psi_c + \beta_{\bar{Y}_{a,bc}} \bar{\phi}_a \bar{\psi}_b\bar{\psi}_c 
= iQ_{ab}\bar{Y}_{c,ad} \bar{\phi}_{c}\bar{\psi}_{d}\bar{\psi}_b - i Q_{ab} Y_{c,bd} \phi_c\psi_a\psi_d \ ,
\end{align}
and we obtain two independent sets of equations
\begin{align}
\beta_Y &= iQ Y \cr
\beta_{\bar{Y}} &= - i \bar{Y}Q \ .
\end{align}
In unitary theories, the second equations would be a consequence of the first by complex conjugation. Finally, the perturbative computation of the $\beta$ functions for the Yukawa couplings remains the same as in unitary theories except that $Y$ and $\bar{Y}$ are independent. Thus, we use 
\begin{align}
\beta_Y &= \gamma Y = \frac{1}{32 \pi^2}(Y\bar{Y})Y + O(Y^5) \cr
\beta_{\bar{Y}} &= \bar{Y}\gamma = \frac{1}{32 \pi^2}\bar{Y}({Y}\bar{Y}) + O(Y^5)
\end{align}
These equations are completely in agreement with \eqref{vseq} and \eqref{vsbq}. 
\section{Vanishing of $S$-function and computation of $\mathcal{B}$ function in SUSY theories}
For our discussions in section 3, it is crucial to realize that the supersymmetric computation of the $\beta$ functions presented there actually gives the gauge invariant $\mathcal{B}$ functions as long as the regularization preserves the manifest supersymmetry (unlike in some non-supersymmetric computations). Since this will be important for the recent debates over the existence of scale invariant but non-conformal field theories  accessible in perturbation theory in $(1+3)$ dimension, we would like to elaborate on it here.\footnote{The author would like to thank H.~Osborn and J.~Polchinski for the correspondence.}

To make this point clear, let us present our current understanding of the models studied in \cite{Fortin:2012cq}\cite{Fortin:2012hn} as well as the earlier ones \cite{Fortin:2011sz}\cite{Fortin:2012ic}  that claim scale invariance without conformal invariance. In these references, they computed the $\beta$ functions of scalars/fermions/gauge coupled system in a fixed renormaization scheme (say, minimal subtraction scheme), and then attempted to solve the equation 
\begin{align}
\beta^I \mathcal{O}_I = \partial^\mu K_{\mu} \ .
\end{align}
They surprisingly found that there exists a non-trivial $K_{\mu}$ beyond three-loop. If the left hand side of this equation were the trace of the energy-momentum tensor, in other words, if the $\beta$ functions computed and used here were the $\mathcal{B}$ functions discussed in Appendix A, then we could declare that we find a scale invariant but non-conformal field theory at the three-loop order. 

In general, this assumption is quite non-trivial. In the flat space-time renormalization scheme, we are not certain whether we are really computing the gauge invariant $\mathcal{B}$ functions or just $\beta$ functions in a particular gauge simply because the counter term $\mathcal{B}^I \mathcal{O}_I$ and $\beta^I \mathcal{O}_I$ both render the loop computation finite, and with no further discussions, we cannot make a distinction.\footnote{After all, flat space Callan-Symanzik equation only tells the relation between beta functions and trace of the energy-momentum tensor up to total divergence terms.}
Thus, we have to recall that the real criterion of the conformal invariance is to find a non-trivial Virial current $J_\mu$ in
\begin{align}
T^{\mu}_{\ \mu} = \mathcal{B}^I \mathcal{O}_I = \beta^I \mathcal{O}_I - \partial^\mu \mathcal{J}_\mu = \partial^\mu J_{\mu} \ . \label{traceem}
\end{align}
We stress that unless we find what is $\partial^\mu \mathcal{J}_\mu$, which amounts to determining the $S$-function that appeared in Appendix A, we cannot assert whether the theory is merely scale invariant or conformal invariant even if we computed the $\beta$ function in a certain gauge.

A further detailed and careful study by the same authors revealed that the partial Virial current $\mathcal{J}_\mu$ or the $S$-function is non-zero within the same regularization scheme at the three-loop order consistently in these models. See Appendix B of the paper \cite{Fortin:2012hn} .
 In addition, they showed that $\mathcal{J}_\mu$ precisely coincides with $K_{\mu}$ along the scale invariant cycles. In the computation of the trace of the energy-momentum tensor \eqref{traceem}, the coincidence leads to the cancellation and makes $T^{\mu}_{\ \mu}$ vanish. 
This concludes that their claimed ``scale invariant but non-conformal cycles" are actually conformal invariant.\footnote{In \cite{Fortin:2012hn} , it was interpreted in a different manner: they claim that this is due to the difference between $\eta$ flow and the ``real" renormalization group flow. Our discussions suggest that $\eta$ flow is the relevant flow for studying conformal invariance. Holographic argument naturally leads to the $\eta$-flow \cite{Nakayama:2012sn}. Of course, if we stick to the gauge that makes  the coupling constant run according to the $\beta$ functions rather than the $\mathcal{B}$ functions, then the flow of the coupling constants certainly looks cyclic. However, physically it is equivalent to a conformal field theory.} 
In this sense, all the power-counting renormalizable and unitary $(1+3)$ dimensional scale invariant field theories found so far in the literature are conformal invariant in complete agreement with the general discussions on perturbative $a$-theorem \cite{Osborn:1991gm}\cite{Luty:2012ww}.

Going back to our supersymmetric models, we have to argue whether the $S$-function is non-zero or not before we make the claim that we really computed the $\mathcal{B}$ functions. Our claim here is that to all orders in perturbation theory, the  supersymmetric preserving regularization must give vanishing $S$-function so that the Virial current multiplet in section 3 is exact. 

The argument is based on the background field renormalization that we typically employ in the supersymmetric field theories. Suppose we treat coupling constants $Y_{ijk}$ as chiral superfields (gauge couplings do not play any role in the following). The relevant Virial current-background current interaction, which is called $({N}_I)_{ab}$ in \cite{Fortin:2012hn}, must be in the K\"ahler renormalization
\begin{align}
\int d^4\theta N^{ij}(Y,\bar{Y}) \Phi^\dagger_i \Phi_j \ .
\end{align}
Indeed, this term must contain the coupling $ (N^{ijk}_{lm} \partial_\mu Y_{ijk} - \bar{N}^{ijk}_{lm} \partial_\mu \bar{Y}_{ijk}) (\phi^{\dagger l} \partial^\mu \phi^m - \phi^{l} \partial^\mu \phi^{\dagger m}) + \text{fermions}$ that is required to make the theory finite if the coupling constant is space-time dependent, which is called $N^I_1$ in reference \cite{Fortin:2012hn}.

The reality of the renormalized action demands $N^{ij}$ to be Hermitian. Unlike the non-supersymmetric examples, the wave-function renormalization and the Virial current-background current interaction are tightly related here. We recall that $N^{ij}$ is nothing but the anomalous dimension matrix used in section 3 to determine the $\beta$ functions when we set $Y$ space-time independent.
Now in order to obtain a non-trivial $S$-function, we have to replace $\partial_\mu Y$ in $N^{ijk}_{lm}$ with $Y$ and read the contribution to the anti-symmetric wave-function renormalization. In our case, we have to study the Hermitian part of
\begin{align}
i \frac{\partial N^{ij}}{\partial Y_{klm}} Y_{klm} - i\frac{\partial N^{ij}}{\partial \bar{Y}_{klm}} \bar{Y}_{klm} \ .
\end{align}
This vanishes to all orders in perturbation theory.\footnote{Alternatively speaking, the manifestly supersymmetric action does not contain any terms that cannot be removed by the wavefunction renormalization even after we make the coupling space-time dependent.} We can directly check that the three-loop results in \cite{Fortin:2012hn}  vanish for supersymmetric field theories. 

We conclude that our formula computes the gauge invariant $\mathcal{B}$ functions to all orders in perturbation theory, and it can be used to study the structure of the Virial multiplet. We emphasize that this alone would not conclude that there is no scale invariant but non-conformal supersymmetric field theories because still we may find a non-trivial Virial multiplet as discussed in section 3. Of course, the perturbative strong $a$-theorem forbids such a possibility in perturbation theory, but the validity of the $a$-theorem is conceptually independent from the vanishing of the $S$-function discussed here.


\begin{thebibliography}{99}
\bibitem{Haag:1974qh}\
  R.~Haag, J.~T.~Lopuszanski and M.~Sohnius,
  Nucl.\ Phys.\  B {\bf 88}, 257 (1975).


\bibitem{Efimov:1970zz} 
  V.~Efimov,
  Phys.\ Lett.\ B {\bf 33}, 563 (1970).

\bibitem{Zamolodchikov:1986gt}
  A.~B.~Zamolodchikov,
  JETP Lett.\  {\bf 43} (1986) 730
  [Pisma Zh.\ Eksp.\ Teor.\ Fiz.\  {\bf 43} (1986) 565].

\bibitem{Polchinski:1987dy}
  J.~Polchinski,
  Nucl.\ Phys.\  B {\bf 303}, 226 (1988).



\bibitem{Mack1}
M. L\"{u}scher and G. Mack,
1976, unpublished;

G. Mack,
in ``NONPERTURBATIVE QUANTUM FIELD THEORY. PROCEEDINGS, NATO
ADVANCED STUDY INSTITUTE, CARGESE, FRANCE, JULY 16-30, 1987''.

\bibitem{Dorigoni:2009ra}
  D.~Dorigoni and S.~Rychkov,
  arXiv:0910.1087 [hep-th].
\bibitem{Jackiw:2011vz} 
  R.~Jackiw and S.~-Y.~Pi,
  J.\ Phys.\ A A {\bf 44}, 223001 (2011)
  [arXiv:1101.4886 [math-ph]].

\bibitem{ElShowk:2011gz} 
  S.~El-Showk, Y.~Nakayama and S.~Rychkov,
  Nucl.\ Phys.\ B {\bf 848}, 578 (2011)
  [arXiv:1101.5385 [hep-th]].

\bibitem{Nakayama:2011tk} 
  Y.~Nakayama,
  arXiv:1109.5883 [hep-th].

\bibitem{Nakayama:2011wq} 
  Y.~Nakayama,
  Mod.\ Phys.\ Lett.\ A {\bf 27}, 1250029 (2012)
  [arXiv:1110.2586 [hep-th]].
\bibitem{Luty:2012ww} 
  M.~A.~Luty, J.~Polchinski and R.~Rattazzi,
  arXiv:1204.5221 [hep-th].

\bibitem{Nakayama:2009fe}
  Y.~Nakayama,
  JHEP {\bf 1001}, 030 (2010)
  [arXiv:0909.4297 [hep-th]].
\cite{Nakayama:2009qu}
\bibitem{Nakayama:2009qu}
  Y.~Nakayama,
  arXiv:0907.0227 [hep-th].
\bibitem{Nakayama:2010wx} 
  Y.~Nakayama,
  Eur.\ Phys.\ J.\ C {\bf 72}, 1870 (2012)
  [arXiv:1009.0491 [hep-th]].

\bibitem{Nakayama:2010zz} 
  Y.~Nakayama,
  Int.\ J.\ Mod.\ Phys.\ A {\bf 25}, 4849 (2010).


\bibitem{Fortin:2012cq} 
  J.~-F.~Fortin, B.~Grinstein and A.~Stergiou,
  arXiv:1206.2921 [hep-th].

\bibitem{Fortin:2012hn} 
  J.~-F.~Fortin, B.~Grinstein and A.~Stergiou,
  arXiv:1208.3674 [hep-th].

\bibitem{Mack:1975je} 
  G.~Mack,
  Commun.\ Math.\ Phys.\  {\bf 55}, 1 (1977).

\bibitem{Minwalla:1997ka} 
  S.~Minwalla,
  Adv.\ Theor.\ Math.\ Phys.\  {\bf 2}, 781 (1998)
  [hep-th/9712074].

\bibitem{Komargodski:2010rb} 
  Z.~Komargodski and N.~Seiberg,
  JHEP {\bf 1007}, 017 (2010)
  [arXiv:1002.2228 [hep-th]].

\bibitem{Dumitrescu:2011iu} 
  T.~T.~Dumitrescu and N.~Seiberg,
  JHEP {\bf 1107}, 095 (2011)
  [arXiv:1106.0031 [hep-th]].


\bibitem{Ferrara:1974pz} 
  S.~Ferrara and B.~Zumino,
  Nucl.\ Phys.\ B {\bf 87}, 207 (1975).

\bibitem{Gates:1981yc} 
  S.~J.~Gates, Jr., M.~T.~Grisaru and W.~Siegel,
  Nucl.\ Phys.\ B {\bf 203}, 189 (1982).


\bibitem{Kuzenko:2010am} 
  S.~M.~Kuzenko,
  JHEP {\bf 1004}, 022 (2010)
  [arXiv:1002.4932 [hep-th]].
\bibitem{Kuzenko:2010ni} 
  S.~M.~Kuzenko,
  Eur.\ Phys.\ J.\ C {\bf 71}, 1513 (2011)
  [arXiv:1008.1877 [hep-th]].
\bibitem{Zheng:2010xx} 
  S.~Zheng and J.~-h.~Huang,
  Class.\ Quant.\ Grav.\  {\bf 28}, 075012 (2011)
  [arXiv:1007.3092 [hep-th]].



\bibitem{Antoniadis:2011gn} 
  I.~Antoniadis and M.~Buican,
  Phys.\ Rev.\ D {\bf 83}, 105011 (2011)
  [arXiv:1102.2294 [hep-th]].


\bibitem{Nakayama:2004vk} 
  Y.~Nakayama,
  Int.\ J.\ Mod.\ Phys.\ A {\bf 19}, 2771 (2004)
  [hep-th/0402009].

\bibitem{Coleman:1970je} 
  S.~R.~Coleman and R.~Jackiw,
  Annals Phys.\  {\bf 67}, 552 (1971).


\bibitem{Grinstein:2008qk} 
  B.~Grinstein, K.~A.~Intriligator and I.~Z.~Rothstein,
  Phys.\ Lett.\ B {\bf 662}, 367 (2008)
  [arXiv:0801.1140 [hep-ph]].


\bibitem{Fortin:2011bm} 
  J.~-F.~Fortin, B.~Grinstein and A.~Stergiou,
  Phys.\ Lett.\ B {\bf 709}, 408 (2012)
  [arXiv:1110.1634 [hep-th]].

\bibitem{Bracken:1982ny} 
  A.~J.~Bracken and B.~Jessup,
  J.\ Math.\ Phys.\  {\bf 23}, 1925 (1982).
\bibitem{Siegel:1988gd} 
  W.~Siegel,
  Int.\ J.\ Mod.\ Phys.\ A {\bf 4}, 2015 (1989).
\bibitem{Dienes:2009td} 
  K.~R.~Dienes and B.~Thomas,
  Phys.\ Rev.\ D {\bf 81}, 065023 (2010)
  [arXiv:0911.0677 [hep-th]].

\bibitem{Buchbinder:2002gh} 
  I.~L.~Buchbinder, S.~J.~Gates, Jr., W.~D.~Linch, III and J.~Phillips,
  Phys.\ Lett.\ B {\bf 535}, 280 (2002)
  [hep-th/0201096].
\bibitem{Gates:2003cz} 
  S.~J.~Gates, Jr.., S.~M.~Kuzenko and J.~Phillips,
  Phys.\ Lett.\ B {\bf 576}, 97 (2003)
  [hep-th/0306288].


\bibitem{Pons:2009nb} 
  J.~M.~Pons,
  J.\ Math.\ Phys.\  {\bf 52}, 012904 (2011)
  [arXiv:0902.4871 [hep-th]].
\bibitem{Buchbinder:1995uq} 
  I.~L.~Buchbinder and S.~M.~Kuzenko,
  Bristol, UK: IOP (1995) 640 p
\bibitem{Siegel:1979ai} 
  W.~Siegel,
  Phys.\ Lett.\ B {\bf 85}, 333 (1979).

\bibitem{Clark:1980dw} 
  T.~E.~Clark, O.~Piguet and K.~Sibold,
  Nucl.\ Phys.\ B {\bf 172}, 201 (1980).

\bibitem{Kogan:1995mr} 
  I.~I.~Kogan, M.~A.~Shifman and A.~I.~Vainshtein,
  Phys.\ Rev.\ D {\bf 53}, 4526 (1996)
  [Erratum-ibid.\ D {\bf 59}, 109903 (1999)]
  [hep-th/9507170].

\bibitem{Leigh:1995ep} 
  R.~G.~Leigh and M.~J.~Strassler,
  Nucl.\ Phys.\ B {\bf 447}, 95 (1995)
  [hep-th/9503121].

\bibitem{Babington:2005vu} 
  J.~Babington and J.~Erdmenger,
  JHEP {\bf 0506}, 004 (2005)
  [hep-th/0502214].

\bibitem{Yonekura:2010mc} 
  K.~Yonekura,
  JHEP {\bf 1009}, 049 (2010)
  [arXiv:1004.1296 [hep-th]].


\bibitem{Konishi:1983hf} 
  K.~Konishi,
  Phys.\ Lett.\ B {\bf 135}, 439 (1984).

\bibitem{Novikov:1983uc} 
  V.~A.~Novikov, M.~A.~Shifman, A.~I.~Vainshtein and V.~I.~Zakharov,
  Nucl.\ Phys.\ B {\bf 229}, 381 (1983).
\bibitem{ArkaniHamed:1997ut} 
  N.~Arkani-Hamed and H.~Murayama,
  Phys.\ Rev.\ D {\bf 57}, 6638 (1998)
  [hep-th/9705189].

\bibitem{Arnold:2012yi} 
  D.~Arnold, J.~-P.~Derendinger and J.~Hartong,
  arXiv:1208.1648 [hep-th].

\bibitem{Osborn:1991gm} 
  H.~Osborn,
  Nucl.\ Phys.\ B {\bf 363}, 486 (1991).

\bibitem{Nakayama:2012sn} 
  Y.~Nakayama,
  arXiv:1203.1068 [hep-th].

\bibitem{Affleck:1983mk} 
  I.~Affleck, M.~Dine and N.~Seiberg,
  Nucl.\ Phys.\ B {\bf 241}, 493 (1984).

\bibitem{Intriligator:2003jj} 
  K.~A.~Intriligator and B.~Wecht,
  Nucl.\ Phys.\ B {\bf 667}, 183 (2003)
  [hep-th/0304128].
\bibitem{Freedman:1998rd} 
  D.~Z.~Freedman and H.~Osborn,
  Phys.\ Lett.\ B {\bf 432}, 353 (1998)
  [hep-th/9804101].


\bibitem{Fortin:2011sz} 
  J.~-F.~Fortin, B.~Grinstein and A.~Stergiou,
  JHEP {\bf 1207}, 025 (2012)
  [arXiv:1107.3840 [hep-th]].

\bibitem{Fortin:2012ic} 
  J.~-F.~Fortin, B.~Grinstein and A.~Stergiou,
  arXiv:1202.4757 [hep-th].


\bibitem{Riva:2005gd} 
  V.~Riva and J.~L.~Cardy,
  Phys.\ Lett.\ B {\bf 622}, 339 (2005)
  [hep-th/0504197].

\bibitem{Ho:2008nr} 
  C.~M.~Ho and Y.~Nakayama,
  JHEP {\bf 0807}, 109 (2008)
  [arXiv:0804.3635 [hep-th]].

\bibitem{Osborn:1998qu} 
  H.~Osborn,
  Annals Phys.\  {\bf 272}, 243 (1999)
  [hep-th/9808041].

\bibitem{Magro:2001aj} 
  M.~Magro, I.~Sachs and S.~Wolf,
  Annals Phys.\  {\bf 298}, 123 (2002)
  [hep-th/0110131].

\bibitem{Grosse:2007au} 
  J.~Grosse,
  Fortsch.\ Phys.\  {\bf 56}, 183 (2008)
  [arXiv:0711.0444 [hep-th]].


\bibitem{Nakayama:2011zw} 
  Y.~Nakayama,
  Mod.\ Phys.\ Lett.\ A {\bf 26}, 2469 (2011)
  [arXiv:1107.2928 [hep-th]].

\bibitem{SN}
Y.~Nakayama, and S.~J.~Rey, unpublished.


\bibitem{Festuccia:2011ws} 
  G.~Festuccia and N.~Seiberg,
  JHEP {\bf 1106}, 114 (2011)
  [arXiv:1105.0689 [hep-th]].

\bibitem{Dumitrescu:2012ha} 
  T.~T.~Dumitrescu, G.~Festuccia and N.~Seiberg,
  arXiv:1205.1115 [hep-th].


\end{thebibliography}
\end{document}